\documentclass
[preprint,prl,superscriptaddress,tightenlines,twocolumn,10pt,showpacs]{revtex4}%
\usepackage{amsfonts}
\usepackage{amsmath}
\usepackage{amssymb}
\usepackage{graphicx}%
\providecommand{\U}[1]{\protect\rule{.1in}{.1in}}
\newtheorem{theorem}{Theorem}
\newtheorem{acknowledgement}[theorem]{Acknowledgement}

\begin{document}
\title{Cavity-meidated collisionless sympathetic cooling of molecules with atoms}
\author{Guangjiong Dong$^{\dag}$,Chang Wang, Weiping Zhang}
\affiliation{State key laboratory of precision spectroscopy, Department of Physics, East
China Normal University, Shanghai 200062, China}


\begin{abstract}
Cooling a range of molecules to ultracold temperatures (%
%TCIMACRO{\TEXTsymbol{<}}%
%BeginExpansion
$<$%
%EndExpansion
1 mK) is a difficult but important challenge in molecular physics and
chemistry. Collective cavity cooling of molecules is a promising method that
doesn't rely on molecular energy level and thus can be applied to all
molecules in principle. However, the initial lack of cold molecules leads to
the difficulty in its experimental implementation. We show that efficient
collective sympathetic cooling of molecules to sub-mK temperatures using a
large ensemble of atoms within a cavity is feasible. This approach is a new
type of sympathetic cooling which does not rely on direct collisions between
atoms and molecules, but utilizes thermalization via their mutual interaction
with a cavity field. Two important mechanisms are identified. This include:
(1) giant enhancement of cavity optical field from the efficient scattering of
the pump light by the atoms; (2) cavity-mediated collective interaction
between the atoms and the molecules. We show an optimal cavity detuning for
maximizing cooling, which is dependent on the atom and molecule numbers. We
determine a threshold for the molecular pump strength and show that it is
independent of molecule number when the number of atoms is much greater than
the molecules. This can be reduced by orders of magnitude when compared to
cavity cooling of single molecular species only. Using this new sympathetic
cavity cooling technique, cooling molecules to sub-mK within a high-Q cavity
could be within reach of experimental demonstration.

\end{abstract}
\maketitle

The creation and application \cite{cm,qm,qc}\ of cold (%
%TCIMACRO{\TEXTsymbol{<}}%
%BeginExpansion
$<$%
%EndExpansion
1K) and ultracold (%
%TCIMACRO{\TEXTsymbol{<}}%
%BeginExpansion
$<$%
%EndExpansion
1 mK) molecules is now being pursued within the molecular physics \cite{bf}
and quantum chemistry \cite{cp,cc}. Cold molecules could be used for molecular
clocks \cite{molcularFountain} covering broader frequency range than atom
clocks. They offer the prospect of testing the time dependence fundamental
constants \cite{ye}. Already cold molecules have been used in precision
molecular spectroscopy with great potential for tests beyond the standard
model \cite{yb}. Ultracold polar molecules with well controlled long-range
anisotropic dipole-dipole interactions are of particular interest for their
application in quantum computation \cite{qc} and in quantum simulation of
exotic strongly-correlated many-body phenomena \cite{cm}. In addition, quantum
matter-wave optics beyond Kerr nonlinear optics has been proposed \cite{dong}.
The cold and ultracold regime offers a new frontier in chemistry where quantum
tunneling dominates in chemical reaction \cite{cc}, opening up a new approach
to the creation of new materials \cite{nm}.

Cold molecules can be generated using many techniques, such as buffer gas
cooling \cite{buf}, electric Stark deceleration \cite{de}, optical lattice
deceleration \cite{d1}, Zeeman deceleration \cite{ze}, and velocity filtering
\cite{vf}. However, ultracold molecules, except for ultracold alkali molecules
via association of ultracold atoms \cite{pa,ma}, are hardly produced. Now,
developing general ways of producing ultracold molecules is a big theoretical
and experimental challenge in ultracold molecule research.

Recently, sympathetic cooling has been attempted for generating ultracold
molecules through collision with pre-cooled atoms \cite{sym}. In this cooling
process, elastic collision can lead to translational cooling of molecules and
even the internal rotation cooling. However, inelastic collision leads to
heating and trap loss of the molecules, thus, a high ratio of elastic
collision rate to inelastic collision rate is critical for efficient
sympathetic cooling. Moreover, the rate of cold chemical reaction which can
quench sympathetic cooling should also be low. Now experimental implementation
of this technique for most of molecules faces the lack of the detailed
collision knowledge.

Doppler cooling and Sisyphus cooling of SrF molecules \cite{dem} have been
demonstrated with multiple lasers for controlling 44 energy levels. However,
such a direct laser cooling is hardly applicable to most of molecules whose
complex energy levels require many repumping lasers required by Doppler
cooling. Collective cavity cooling \cite{ri,r,temperature,lev,ri1,lu,lu1},
which uses cavity leakage to dissipate kinetic energy, is a candidate for
cooling molecules to sub-mK regime. To avoid direct using molecular energy
levels, a far-off resonant pump laser is required. Consequently, it is not
capable of generating the sufficiently strong dispersive coupling to the
cavity field, leading to slow cavity cooling and much higher pump thresholds
for cooling. To overcome these difficulties, a huge number ($\sim10^{9}$
\cite{lu1}) of cold molecules are needed. Unfortunately, it is difficult for
current methods \cite{buf,de,d1,ze,vf} to produce the large number of cold
molecules required.

In this letter, we show that collective sympathetic cooling of molecules with
atoms mediated by an optical cavity is capable of rapidly cooling essentially
any molecular species. This cooling uses the effective enhancement of the
cavity optical field by the large number of ultra cold atoms that can be
placed in the cavity which cools the much smaller number of molecules. We find
that when both atoms and molecules within a high-Q cavity are side pumped with
a strong far-off resonant light, the cooling rate of molecules can be
significantly enhanced by increasing the number of atoms,while reducing the
pump laser power by orders of magnitude. Our new cooling scheme does not need
a large number of cold molecules and does not require knowledge of the
complicated, typically inelastic interactions of conventional sympathetic
cooling schemes. These simplifications put it firmly within reach of
experimental realization.%
%TCIMACRO{\FRAME{ftbpFU}{1.8896in}{0.9885in}{0pt}{\Qcb{Atoms and molecules are
%loaded in the cavity, and side pumped by a far-off resonant light with
%frequency $\omega_{p}$. Cavity optical field leaks from the cavity with a rate
%$\kappa$.}}{}{Figure}{\special{ language "Scientific Word";  type "GRAPHIC";
%maintain-aspect-ratio TRUE;  display "USEDEF";  valid_file "T";
%width 1.8896in;  height 0.9885in;  depth 0pt;  original-width 7.5316in;
%original-height 3.9271in;  cropleft "0";  croptop "1";  cropright "1";
%cropbottom "0";  tempfilename 'M2VE5903.wmf';tempfile-properties "XPR";}} }%
%BeginExpansion
\begin{figure}
[ptb]
\begin{center}
\includegraphics[
natheight=3.927100in,
natwidth=7.531600in,
height=0.9885in,
width=1.8896in
]%
{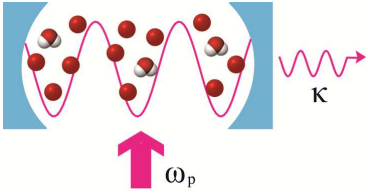}%
\caption{Atoms and molecules are loaded in the cavity, and side pumped by a
far-off resonant light with frequency $\omega_{p}$. Cavity optical field leaks
from the cavity with a rate $\kappa$.}%
\end{center}
\end{figure}
%EndExpansion

Our scheme is shown in Fig. 1. Atoms of mass $m_{a}$ with spontaneous emission
rate $\gamma_{a}$, and molecules of mass $m_{m}$ with spontaneous emission
rate $\gamma_{m}$ within a cavity with a leakage rate $\kappa$ and a resonant
frequency $\omega_{c}$ are side pumped by an optical field with frequency
$\omega_{p}$. In our following analysis, we assume that the pump frequency is
far-off resonant from any transition of the atoms and molecules, for the
purpose of developing a technique without using the detailed knowledge of
molecular energy levels. \ For simplifying our analysis, we make two other
assumptions: (1) the transverse motion of the atoms and molecules are frozen,
so that we only need to investigate their motion along the cavity axis; (2)
any potential collision between atoms and molecules has been neglected. With
these assumptions and extension of the cavity cooling theory for atoms
\cite{ri,r,temperature}, the cavity optical field $\alpha$ and the
center-of-mass motion for the atoms and the molecules are described by the
following equations:%
\begin{align}
\frac{d\alpha}{dt}  &  =\left(  i\Delta_{C}-\kappa\right)  \alpha-\sum
_{j}\left[  \left(  iU_{0}^{a}+\Gamma_{0}^{a}\right)  \cos^{2}\left(
\tilde{z}_{j}^{a}\right)  \alpha+\eta_{eff}^{a}\cos\left(  \tilde{z}_{j}%
^{a}\right)  \right] \nonumber\\
&  -\sum_{l}\left[  \left(  iU_{0}^{m}+\Gamma_{0}^{m}\right)  \cos^{2}\left(
\tilde{z}_{l}^{m}\right)  \alpha+\eta_{eff}^{m}\cos\left(  \tilde{z}_{l}%
^{m}\right)  \right]  +\xi_{\alpha} \label{1}%
\end{align}%
\begin{equation}
\frac{d\tilde{p}_{j}^{a}}{dt}=\mathfrak{N}U_{0}^{a}\sin\left(  2\tilde{z}%
_{j}^{a}\right)  +2\operatorname{Im}(\eta_{eff}^{a}\alpha^{\ast})\sin\left(
\tilde{z}_{j}^{a}\right)  +\xi_{j}^{a} \label{2}%
\end{equation}%
\begin{equation}
\frac{dz_{j}^{a}}{dt}=\frac{p_{j}^{a}}{m_{a}} \label{3}%
\end{equation}%
\begin{equation}
\frac{d\tilde{p}_{l}^{m}}{dt}=\mathfrak{N}U_{0}^{m}\sin\left(  2\tilde{z}%
_{l}^{m}\right)  +2\operatorname{Im}\left(  \eta_{eff}^{m}\alpha^{\ast
}\right)  \sin\left(  \tilde{z}_{l}^{m}\right)  +\xi_{l}^{m} \label{4}%
\end{equation}%
\begin{equation}
\frac{dz_{l}^{m}}{dt}=\frac{p_{l}^{m}}{m_{m}} \label{5}%
\end{equation}
in which $\Delta_{\text{C}}=\omega_{\text{P}}-\omega_{\text{C}}$ , $p_{j}^{a}$
and $p_{l}^{m}$ are respectively the momentum for the \textit{j}-th atoms and
the \textit{l}-th molecules, and their corresponding positions are denoted
with $z_{j}^{a}$ and $z_{l}^{m}$. Their momentums and coordinates are scaled
by $\tilde{p}_{j}^{a}=p_{j}^{a}/(\hbar k_{c})$, $\tilde{p}_{l}^{m}=p_{j}%
^{m}/(\hbar k_{c})$, $\tilde{z}_{j}^{a}=k_{c}z_{j}^{a}$, $\tilde{z}_{l}%
^{m}=k_{c}z_{l}^{m}$ ($k_{c}$ is the wave number of the cavity optical field).
Parameter $\mathfrak{N}$ is given by $\mathfrak{N}=\left\vert \alpha
\right\vert ^{2}-1/2$. Parameters $U_{0}$ and $\Gamma_{0}$ respectively denote
the dispersive and absorption effect, and $\eta_{eff}$ is the effective
pumping strength. For the atoms and the molecules respectively with the
detuning $\Delta_{a}$ and $\Delta_{m}$ under the pump optical field, these
parameters are given by $U_{0}^{a}=g_{a}^{2}\Delta_{a}/\left(  \Delta_{a}%
^{2}+\gamma_{a}^{2}\right)  $, $U_{0}^{m}=g_{m}^{2}\Delta_{m}/\left(
\Delta_{m}^{2}+\gamma_{m}^{2}\right)  $, $\Gamma_{0}^{a}=g_{a}^{2}\gamma
_{a}/\left(  \Delta_{a}^{2}+\gamma_{a}^{2}\right)  $, $\Gamma_{0}^{m}%
=g_{m}^{2}\gamma_{m}/\left(  \Delta_{m}^{2}+\gamma_{m}^{2}\right)  $,
$\eta_{eff}^{a}=\eta_{a}g_{a}/\left(  -i\Delta_{a}+\gamma_{a}\right)  $ and
$\eta_{eff}^{m}=\eta_{m}g_{m}/\left(  -i\Delta_{m}+\gamma_{m}\right)  $ with
atom/molecule-cavity coupling $g_{a/m}$ and the pump strength $\eta_{a/m}$
\cite{note}. The noise terms can be decomposed into $\xi_{\alpha}=\xi_{\alpha
}^{0}+\xi_{\alpha}^{1}$, $\xi_{j}^{a}=\xi_{j}^{a,0}+\xi_{j}^{a,1}$, $\xi
_{l}^{m}=\xi_{l}^{m,0}+\xi_{l}^{m,1}$ which are given by%
\begin{equation}
\left\langle \left\vert \xi_{\alpha}^{0}\right\vert ^{2}\right\rangle
=\kappa+\frac{\Gamma_{0}^{(a)}}{2}\sum_{j}\cos^{2}(\tilde{z}_{j}^{a}%
)+\frac{\Gamma_{0}^{(m)}}{2}\sum_{i}\cos^{2}(\tilde{z}_{i}^{m}), \label{6}%
\end{equation}%
\begin{equation}
\left\langle \left\vert \xi_{j}^{a,0}\right\vert ^{2}\right\rangle
=2\overline{u^{2}}\Gamma_{0}^{a}\left[  \mathfrak{N}\cos^{2}(\tilde{z}_{j}%
^{a})+2\tilde{\eta}_{a}\alpha_{r}\cos(\tilde{z}_{j}^{a})+\tilde{\eta}_{a}%
{}^{2}\right]  , \label{7}%
\end{equation}%
\begin{equation}
\left\langle \left\vert \xi_{l}^{m,0}\right\vert ^{2}\right\rangle
=2\overline{u^{2}}\Gamma_{0}^{m}\left[  \mathfrak{N}\cos^{2}(\tilde{z}_{l}%
^{m})+2\tilde{\eta}_{m}\alpha_{r}\cos(\tilde{z}_{l}^{m})+\tilde{\eta}_{m}%
{}^{2}\right]  , \label{8}%
\end{equation}%
\begin{equation}
\xi_{\alpha}^{1}=\frac{i\alpha}{\left\vert \alpha\right\vert }\left[
\sqrt{\frac{\Gamma_{0}^{a}}{2}}\sum_{j}\cos(\tilde{z}_{j}^{a})\zeta_{j}%
^{a}+\sqrt{\frac{\Gamma_{0}^{m}}{2}}\sum_{l}\cos\left(  \tilde{z}_{l}%
^{(m)}\right)  \zeta_{l}^{(m)}\right]  , \label{9}%
\end{equation}%
\begin{equation}
\xi_{j}^{a,1}=-\sqrt{2\Gamma_{0}^{a}}\left\vert \alpha\right\vert \sin
(\tilde{z}_{j}^{a})\zeta_{j}^{(a)}, \label{10}%
\end{equation}%
\begin{equation}
\xi_{l}^{m,1}=-\sqrt{2\Gamma_{0}^{m}}\left\vert \alpha\right\vert \sin
(\tilde{z}_{i}^{m})\zeta_{i}^{m}, \label{11}%
\end{equation}
in which $\tilde{\eta}_{a}=\eta_{a}/g_{a}$, $\tilde{\eta}_{m}=\eta_{m}/g_{m}$,
and $\alpha_{r}=\operatorname{Re}(\alpha)$, $k_{c}\sqrt{\overline{u^{2}}}$
represents the mean projection of the momentum recoil along the cavity axis
due to spontaneous emission.

Before numerically solving above equations of motion, we first qualitatively
analyze the possible mechanisms for the collisionless sympathetic cooling of
molecules with the atoms. First, this collisionless sympathetic cooling uses
the giant enhancement of the cavity optical field from the efficient
scattering of the pump light by the atoms. As shown in Eq. (1), the cavity
optical field is induced by the scattered light from both the atoms and the
molecules. When the molecules are pumped by a very far-off resonant light
(typically, $\Delta_{m}\sim10^{15}$ Hz), the scattering light from a small
number of molecules is negligibly weak; however the cavity optical field can
be effectively enhanced by adding the atoms with a off-resonant detuning
($\Delta_{a}\sim10$ GHz), and thus cooling efficiency of the molecules can be
improved with the assistance of the atoms. Second, this collisionless
sympathetic cooling profits from cavity-mediated collective interaction
between the atoms and molecules. Eqs. (2) and (4) show that the optical forces
on an atom or a molecule are functions of the cavity optical field. Whereas,
the cavity optical field, as shown in Eq. (1), is influenced by the motion of
all the atoms and molecules, thus the optical force exerted on a single atom
or a single molecule is contributed from the influence from all other atoms
and molecules, which means that cavity photon-mediated interaction between
atoms and molecules exists \cite{pi}, leading to possibility of sympathetic
cooling of molecules with fast cooled atoms. It is hard for us to single out
familiar two--body interaction between one single atom and one single
molecule, but collective interaction between atoms and molecules exists. So, a
large number of atoms and molecules can enhance the collective interaction and
further speed up the sympathetic cooling of molecules.

We solved the above equations of motion for Rb atoms and CN molecules at
initial temperatures 10 mK with $\Delta_{a}=-20$ GHz, $\Delta_{m}%
=2.45\times10^{15\text{ }}$Hz, cavity detuning $\Delta_{c}=-\kappa+U_{0}%
^{a}N_{a}+U_{0}^{m}N_{m}$ ($N_{a}$ and $N_{m}$ are the total number of the
atoms and the molecules respectively), and $\kappa=20$ MHz. Figure 1 shows the
time evolution of temperatures $T$ of the atoms and the molecules at a pump
strength $\eta_{a}=5000\kappa$ for three cases: (1), cavity cooling of atoms
only ($N_{m}=0$, $N_{a}=10^{6}$, black line is for the atomic temperature);
(2) cavity cooling of molecules only ($N_{m}=10^{4}$, $N_{a}=0$, red line is
for the molecular temperature); (3) cavity cooling of atoms and molecules
mixture ($N_{m}=10^{4}$, $N_{a}=10^{6}$, the blue line is for the atomic
temperature, and the yellow line is for molecule temperature). As shown in
Fig. 1, cavity cooling efficiency of molecules is much enhanced by adding a
large number of atoms into the cavity. The atoms are cooled faster than the
molecules, so the relatively cooler atoms can cool the relatively hotter
molecules through cavity-mediated collective interaction between atoms and
molecules. Consequently, comparing with black line (only atoms), cooling of
the atoms is slowed down when the molecules is added. This slowing down is a
signal of collisionless sympathetic cavity cooling of molecules with atoms.%
%TCIMACRO{\FRAME{ftbpFU}{3.2558in}{2.7345in}{0pt}{\Qcb{Times evolution of the
%temperatures for the atom ensemble and the molecule ensemble for three cases:
%(1) only 10$^{6}$ atoms are in the cavity (black line); (2) only 10$^{4}$
%molecules are in the cavity (red line); (3)10$^{6}$ atoms and 10$^{4}$
%molecules are in the same cavity (yellow line for molecules, and blue line
%\ for atoms).}}{}{newcase.eps}{\special{ language "Scientific Word";
%type "GRAPHIC";  maintain-aspect-ratio TRUE;  display "USEDEF";
%valid_file "F";  width 3.2558in;  height 2.7345in;  depth 0pt;
%original-width 11.6949in;  original-height 8.2538in;  cropleft "0";
%filename 'newcase.eps';file-properties "XNPEU";}} }%
%BeginExpansion
\begin{figure}
[ptb]
\begin{center}
\includegraphics[
height=2.7345in,
width=3.2558in
]%
{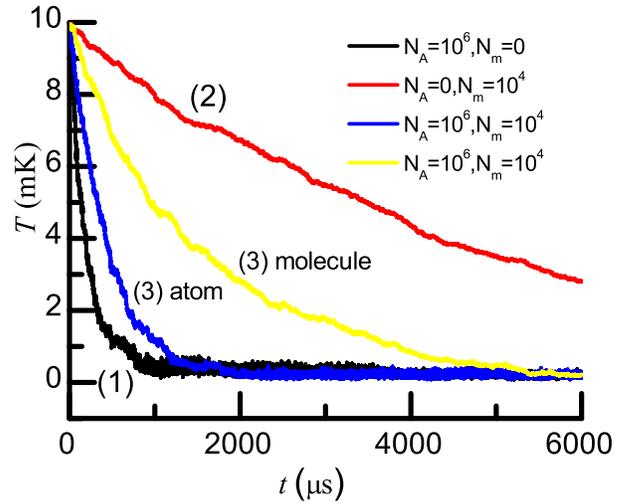}%
\caption{Times evolution of the temperatures for the atom ensemble and the
molecule ensemble for three cases: (1) only 10$^{6}$ atoms are in the cavity
(black line); (2) only 10$^{4}$ molecules are in the cavity (red line);
(3)10$^{6}$ atoms and 10$^{4}$ molecules are in the same cavity (yellow line
for molecules, and blue line \ for atoms).}%
\end{center}
\end{figure}
%EndExpansion

We further study the relation of sympathetic cavity cooling efficiency to the
pumping power. Figure 3 shows the time evolution of temperature for 10$^{4}$
molecules assisted by 10$^{6}$ atoms for pumping strength $\eta_{a}=500\kappa$
(red line), and 5000$\kappa$ (black line). Figure 3 shows that cooling time
increases with the decreasing of the pump power. We can imagine that the
sympathetic cavity cooling of molecules with atoms has minimum requirement for
the pump laser power.%
%TCIMACRO{\FRAME{ftbpFU}{3.3481in}{2.0863in}{0pt}{\Qcb{Time evolution of the
%molecule temperature when $10^{6}$ atoms and $N_{m}=10^{4}$ molecules are in
%the cavity for different pump strengths $\eta_{a}=5000\kappa$ (black) and
%500$\kappa$ (red) }}{}{eta1.eps}{\special{ language "Scientific Word";
%type "GRAPHIC";  maintain-aspect-ratio TRUE;  display "USEDEF";
%valid_file "F";  width 3.3481in;  height 2.0863in;  depth 0pt;
%original-width 11.6949in;  original-height 8.2538in;  cropleft "0";
%filename 'eta1.eps';file-properties "XNPEU";}} }%
%BeginExpansion
\begin{figure}
[ptb]
\begin{center}
\includegraphics[
height=2.0863in,
width=3.3481in
]%
{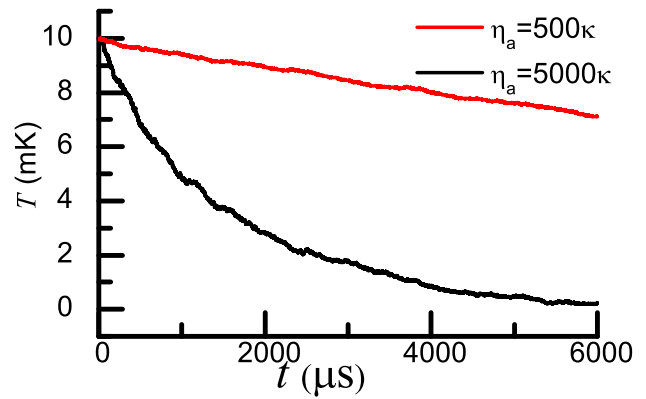}%
\caption{Time evolution of the molecule temperature when $10^{6}$ atoms and
$N_{m}=10^{4}$ molecules are in the cavity for different pump strengths
$\eta_{a}=5000\kappa$ (black) and 500$\kappa$ (red) }%
\end{center}
\end{figure}
%EndExpansion
\ 

We now discuss the threshold of the pump laser power. Similar to cavity
cooling of single species, sympathetic cavity cooling of molecules relies on
self-organization of molecules and atoms in the cavity. The atoms and the
molecules are initially distributed uniformly in the cavity with statistic
fluctuation. When the pump laser is turned on, the cavity optical field starts
to build up through light scattering from the molecules and atoms. The
intra-cavity field and the scattering optical field will lead to localization
of molecules and atoms in positions around the position $k_{c}x=2n\pi$ and
($2n+1)\pi$ ($n$ is an integer). Such a self-organization requires the
intra-cavity field is strong enough to produce a deep potential well to
suppress the thermal motion. Since the dispersion couplings $U_{0}^{a}$ and
$U_{0}^{m}$ are weak in the far-off resonant optical field, the potential well
depths for the atoms and molecules are respectively approximated by $\ $%
\begin{equation}
U_{a/m}\approx2\hbar\eta_{a/m}g_{a/m}\alpha_{r}/\Delta_{a/m}. \label{q}%
\end{equation}
The real part of $\alpha$, $\alpha_{r}$, can be estimated by the steady
solution of the Eq. (1), which is given by
\begin{equation}
\alpha_{r}=\frac{2(\frac{\eta_{m}g_{m}}{\Delta_{m}}\delta N_{m}+\frac{\eta
_{a}g_{a}}{\Delta_{a}}\delta N_{a})[\Delta_{c}-\frac{U_{0}^{a}N_{a}+U_{0}%
^{m}N_{m}}{2}]}{\pi\left\{  \left[  \Delta_{c}-\frac{U_{0}^{a}N_{a}+U_{0}%
^{m}N_{m}}{2}\right]  ^{2}+\left[  \kappa+\frac{\Gamma_{0}^{a}N_{a}+\Gamma
_{0}^{m}N_{m}}{2}\right]  ^{2}\right\}  }. \label{a}%
\end{equation}
In deducing Eq. (\ref{a}), we have used $\cos^{2}\left(  k_{C}z_{j}%
^{(a)}\right)  \approx N_{a}/2,\cos^{2}\left(  k_{C}z_{i}^{(a)}\right)
\approx N_{m}/2$, $\cos\left(  k_{C}z_{j}^{(a)}\right)  \approx2N_{a}/\pi$,
and $\cos\left(  k_{C}z_{i}^{(a)}\right)  \approx2N_{m}/\pi$ in the vicinity
of threshold \cite{temperature,lu1}.

When
\begin{equation}
\Delta_{c}=\pm\left[  \kappa+\frac{\Gamma_{0}^{a}N_{a}+\Gamma_{0}^{m}N_{m}}%
{2}\right]  +U_{0}^{a}N_{a}+U_{0}^{m}N_{m}, \label{opti}%
\end{equation}
$\alpha_{r}(\Delta_{c})$ gets its maximum value, and thus achieving the
deepest potential well depths for $U_{a}$ and $U_{m}$. So, in the following,
we keep the optimum cavity detuning given by Eq. (\ref{opti}), and now%

\begin{equation}
\alpha_{r}=\frac{\left(  \eta_{m}g_{m}/\Delta_{m}\delta N_{m}+\eta_{a}%
g_{a}/\Delta_{a}\delta N_{a}\right)  }{\pi\left(  \kappa+\frac{\Gamma_{0}%
^{a}N_{a}+\Gamma_{0}^{m}N_{m}}{2}\right)  }. \label{ar}%
\end{equation}

Equation (\ref{ar}) shows that when $\Delta_{a}\ll\Delta_{m}$ and $N_{m}\ll
N_{A}$, $\eta_{m}g_{m}/\Delta_{m}\delta N_{m}\ll\eta_{a}g_{a}/\Delta_{a}\delta
N_{a}$, the cavity optical field dominantely comes from the scattering of the
pump light by the atoms. In this situation, using the self-organization
requirement $U_{m}\geq k_{B}T_{f}$ \cite{lu1}, in which $T_{f}=2\hbar
\kappa/k_{B}$ is the finally achieved cooling temperature \cite{temperature},
we get a formula for the threshold for the molecular pump strength,%
\begin{equation}
\eta_{m}^{th}=\sqrt{\frac{\pi}{2Rg_{m}g_{a}}\frac{\kappa^{2}\Delta_{m}%
\Delta_{a}}{\delta N_{a}}}. \label{etam}%
\end{equation}
in which $R=g_{a}/g_{m}$ \cite{n1}.

If only molecules are cooled within the cavity, the threshold $\eta_{m}%
^{th,s}=\sqrt{\pi\kappa^{2}\Delta_{m}^{2}/(2g_{m}^{2}\delta N_{m})}$. The
ratio of $\eta_{m}^{th}$ to $\eta_{m}^{th,s}$ is
\begin{equation}
\eta_{m}^{th}/\eta_{m}^{th,s}=\sqrt{\frac{g_{m}}{Rg_{a}}\frac{\Delta_{a}%
}{\Delta_{m}}\frac{\delta N_{m}}{\delta N_{a}}}. \label{r}%
\end{equation}
Using typical values $\Delta_{m}\sim10^{15}$ Hz, $\Delta_{a}\sim10^{10}$ Hz,
$R\approx$14.7 and $g_{a}/g_{m}$=4.4 in our numerical calculation, even when
$\delta N_{a}=\delta N_{m}$, $\eta_{m}^{th}/$ $\eta_{m}^{th,s}\approx
3.93\times10^{-4}$. Since $\delta N_{a/m}\sim\sqrt{N_{a/m}}$, this ratio can
be further lowered by increasing the ratio $N_{A}/N_{m}$. Thus, the threshold
for molecular pump strength $\eta_{m}^{th}$ with the assistance of atoms can
be reduced by at least four orders of magnitude, compared with $\eta
_{m}^{th,s}$ without using atoms. Physically, this greatly reduced threshold
arises from the much enhanced cavity optical field from the scattering of the
pump light by the atoms.

Equation (\ref{etam}) shows several ways to reduce the threshold of the
pumping power for molecules: reducing the atom detuning and increasing the
atom number. Also, using a cavity with a very low leakage rate $\kappa$ is
useful for reducing the threshold.

Similarly, the threshold for atomic pump strength is $\eta_{a}^{th}=\sqrt
{\pi\kappa^{2}\Delta_{a}^{2}/(2g_{a}^{2}\delta N_{a})}$ which is in general
much smaller than that for cooling molecules when $\Delta_{a}\ll\Delta_{m}$.
Thus, we observe atoms are always cooled before molecules, as shown in Fig 2,
and the cooled atoms can be further used to collective sympathetic cooling of
molecules through cavity-mediated interaction between the atoms and the molecules.

In summary, we have shown that the efficiency of collective cavity cooling of
molecules with the assistance of atoms within the same cavity can be greatly
improved. We have obtained a formula determining the optimum cavity detuning
for efficient cooling. We also have deduced an analytical formula for the pump
power threshold for molecular cavity cooling, and found that the threshold,
independent of the molecule number, can be reduced by several orders of
magnitude. Two mechanisms contributed to the enhanced cavity cooling of
molecules are (1) the enhancement of cavity optical field from the efficient
scattering of the pump light by the atoms; (2) cavity-mediated collective
interaction between atoms and molecules. Using the current cavity cooling
scheme, cavity cooling of even a small number of cold molecules down to sub-mK
temperatures is experiment feasible.

Our scheme presents a new kind of sympathetic cooling, which does not use
direct collision between atoms and molecule. However, when a larger ensemble
of atoms is loaded into the cavity with the molecules, the direct collision
between the atoms and molecules, which has been neglected in our analysis,
will come to play. In general, elastic collision will enhance the cavity
cooling of molecules. However, the role of inelastic collision is unclear.
Moreover, chemical reaction within the cavity could also be induced. To avoid
these direct collision effects, we can use a potential barrier to separate
atoms and molecules in the cavity.

Our results will have implications for a number of future theoretical and
experimental issues, such as single molecule manipulation within a cavity,
cavity-mediated cold collision and cold chemistry. Especially, with the cavity
leakage rate is further reduced in future, ultracold dense degenerate
molecular gases wold be produced within a cavity directly, getting rid of
evaporative cooling.

\begin{acknowledgments}
\begin{acknowledgement}
In the preparation of this manuscript, from private communication with Prof.
H. Ritsch, we learn that a similar idea has appeared in Ref. \cite{ax}. This
work was supported by the National Basic Research Program of China (973
Program) under Grant Nos. 2011CB921604 and 2011CB921602, and the National
Natural Science Foundation of China under Grants Nos. 10874045, 11034002 and 10828408.
\end{acknowledgement}
\end{acknowledgments}

$\dag$ Corresponding authors. Emails: dong.guangjiong@gmail.com


\begin{thebibliography}{99}
\bibitem {note}the pump strength used in Refs. \cite{ri,r} is actually the
Rambi frequency for the atom/molecule under the pump field.
\bibitem {n1}The ratio R is ratio of the atomic Rabi frequency to the
molecular ratio \cite{note}, and thus is a constantt value,
\end{thebibliography}
\end{document}